\documentclass[conference, 10pt]{IEEEtran}
\IEEEoverridecommandlockouts
\usepackage{cite}
\usepackage{hyperref}
\usepackage{amsmath,amssymb,amsfonts}
\usepackage{algorithmic}
\usepackage{graphicx}
\usepackage{textcomp}
\usepackage{xcolor}
\usepackage{bm}
\def\BibTeX{{\rm B\kern-.05em{\sc i\kern-.025em b}\kern-.08em
    T\kern-.1667em\lower.7ex\hbox{E}\kern-.125emX}}

\def\R{\mathbb{R}}
\DeclareRobustCommand{\bigO}{%
  \text{\usefont{OMS}{cmsy}{m}{n}O}%
}

\begin{document}

\title{Quantum Circuit Optimization through Iteratively Pre-Conditioned Gradient Descent
\thanks{Supported by the National Quantum Lab, jointly between the University of Maryland and IonQ Inc. *These two authors contributed equally.}
}

\author{
\IEEEauthorblockN{Dhruv Srinivasan*}
\IEEEauthorblockA{\textit{Department of Mechanical Engineering \& Physics} \\
\textit{University of Maryland}\\
College Park, MD, USA \\
dhruvs@umd.edu}

\and

\IEEEauthorblockN{Kushal Chakrabarti*}
\IEEEauthorblockA{\textit{Division of Data \& Decision Sciences} \\
\textit{Tata Consultancy Services Research}\\
Mumbai, India \\
chakrabarti.k@tcs.com}

\and

\IEEEauthorblockN{Nikhil Chopra}
\IEEEauthorblockA{\textit{Department of Mechanical Engineering} \\
\textit{University of Maryland}\\
College Park, MD, USA\\
nchopra@umd.edu}

\and

\IEEEauthorblockN{Avik Dutt}
\IEEEauthorblockA{\textit{Department of Mechanical Engineering and IPST} \\
\textit{QLab Fellow, National Quantum Laboratory at Maryland} \\
\textit{University of Maryland}\\
College Park, MD, USA\\
avikdutt@umd.edu}
}

\maketitle

\begin{abstract}
Gate-based quantum algorithms are typically implemented by circuits consisting of many single-qubit and multi-qubit gates operating on a quantum input state, followed by measurements of the output state. For certain quantum subroutines, such as initial state preparation and quantum Fourier transforms, explicit decompositions of the circuit in terms of single-qubit and two-qubit maximally entangling gates may exist. However, they often lead to large-depth circuits that are challenging for noisy intermediate-scale quantum (NISQ) hardware. Additionally, exact decompositions might only exist for some modular quantum circuits. Therefore, it is essential to find gate combinations that approximate these circuits to high fidelity  with potentially low depth. Gradient-based optimization has been used to find such approximate decompositions. Still, these traditional optimizers often run into problems of slow convergence requiring many iterations, and performing poorly in the presence of noise, a factor that is especially relevant for NISQ hardware. Here we present iteratively preconditioned gradient descent (IPG) for optimizing quantum circuits and demonstrate performance speedups for state preparation and implementation of quantum algorithmic subroutines. IPG is a noise-resilient, higher-order algorithm that has shown promising gains in convergence speed for classical optimizations, converging locally at a linear rate for convex problems and superlinearly when the solution is unique. Specifically, we show an improvement in fidelity by a factor of $10^4$ for preparing a 4-qubit W state and a maximally entangled 5-qubit GHZ state with compared to other commonly used classical optimizers tuning the same ansatz. We also show performance gains for optimizing a full quantum circuit unitary using IPG, and report on results of running such an optimized quantum Fourier transform (QFT) circuit on IonQ's quantum processing unit (QPU) Aria. Such faster convergence with promise for noise-resilience could provide advantages for quantum algorithms on NISQ hardware, especially since the cost of running each iteration on a quantum computer is substantially higher than the classical optimizer step.

\end{abstract}


\section{Introduction}
Quantum computing promises exponential speedups in certain tasks compared to their classical counterparts by harnessing principles of superposition, quantum interference, and entanglement \cite{nielsen_quantum_2000}. Similarly, digital quantum simulation promises to emulate and predict properties of systems that are classically intractable, enabling quantum chemistry and materials science advances \cite{feynman_simulating_1982, cirac_goals_2012, lloyd_universal_1996}. The most traditional route to quantum computing and digital quantum simulation uses circuits composed of single-qubit and two-qubit gates operating on an input state that encodes the quantum information \cite{nielsen_quantum_2000}. This is referred to as the gate-based model of quantum computing. 

Since fault-tolerant quantum computation operating on error-corrected logical qubits is currently challenging to scale to a large number of qubits, it is essential to find optimal representations of quantum algorithms in terms of low-depth circuits with fewer gate counts than explicit decompositions. Currently, explicit decompositions are efficient (i.e. short-depth circuits exist) only for certain classes of states, and use multicontrolled unitaries that are not native to most qubit hardware platforms \cite{bartschi_short-depth_2022, kaye_quantum_2001, holmes_efficient_2020}; but arbitrary quantum circuits remain challenging to decompose into compact, finite gate sets. Moreover, finding an exact decomposition of the unitary corresponding to a quantum algorithm usually requires concerted manual efforts when complex multistep algorithms operating on many qubits are involved. Reducing gate counts and circuit depths becomes especially important in the ongoing noisy intermediate-scale quantum (NISQ) era \cite{preskill_quantum_2018}, where high gate counts mean significant accumulated errors in the computation. A natural route to circumvent these challenges would be to use traditional optimization techniques for seeking low-depth circuits that approximate the desired unitary to high accuracy (fidelity).

Several machine learning techniques have been used to optimize quantum circuits for specific hardware architectures by various groups. Examples include the use of deep reinforcement learning by Max-Planck/Google \cite{fosel_quantum_2021, ostaszewski_reinforcement_2021}, optimizer-agnostic quantum neural networks by Google \cite{verdon_learning_2019}, genetic algorithms combined with symbolic algebra \cite{wong_algorithm_2012}, and the use of Gaussian processes by LBNL \cite{muller_accelerating_2022}. Hardware architecture-specific versions of approximate quantum circuits have also been proposed using photonic qubits mediated by natural or artificial atoms \cite{bartlett_universal_2020, shi_quantum_2022, bartlett_deterministic_2021}, or for measurement-based quantum computation \cite{da_silva_global_2013}. However, to the best of our knowledge, these works used low-order techniques such as conventional gradient descent \cite{bartlett_universal_2020} or Adam or avoided gradients altogether using e.g., image filtering to evade local minima, as presented in IEEE QCE 2022 \cite{muller_accelerating_2022}. In classical optimization scenarios, first-order gradient descent techniques have been shown to suffer from slow convergence and high sensitivity to noise, particularly when the solution space is non-unique \cite{chakrabarti_accelerating_2021, chakrabarti_accelerating_2021-1}.

Given the many non-unique solutions that can approximate a desired quantum circuit with high fidelity, it is imperative to consider higher-order gradient-descent techniques that show faster convergence and exhibit noise resilience. In this paper, we propose and numerically illustrate the use of a newly developed optimization technique -- that of iteratively preconditioned gradient descent (IPG) -- for optimizing quantum circuits without needing any ancilla qubits. We show its superior convergence to higher fidelity in substantially fewer iterations than conventional gradient descent algorithms. Our work harnesses the concept of differentiable quantum programming to accurately calculate gradients and Hessians that feed into the IPG optimizer. The importance of our contribution is underscored by the fact that each ``pass" through a quantum circuit is more expensive in terms of time  and hardware costs than the cost of more compute power on the classical optimizer, and hence techniques such as IPG that converge faster with a fewer number of iterations could have a significant impact in the NISQ era.

The rest of the paper is organized as follows. In Section II, we provide an overview of the IPG method. In Section III, we discuss our application of the IPG method to the problem of quantum circuit optimization, whose results are presented in Section IV. We conclude with a discussion in Section V.

\section{Background on the IPG method}
This section introduces the Iteratively Preconditioning Gradient (IPG) descent methodology, which aims to compute a minimum point of the {\em cost function} $f:\R^d \to \R$. Formally, the goal is to compute a parameter vector $x_* \in \R^d$ such that
\begin{align}
    x_* \in X_* = \arg \min_{x \in \R^d} f(x). \label{eqn:opt_1}
\end{align}
When the cost function is non-convex, instead of searching a global optimal solution, a more meaningful and attainable goal is to find a {\em stationary point} of the cost function $f$, defined as $x_{st} \in X_{st} = \{x\in \R^d : \nabla f(x) = 0_d\}$,
where $\nabla f(x) \in \R^d$ denotes the gradient of $f$ at $x \in \R^d$.

Built upon the prototypical gradient-descent (GD) algorithm~\cite{bertsekas1989parallel}, several {\em accelerated} and {\em adaptive} gradient algorithms have been proposed for solving~\eqref{eqn:opt_1}~\cite{nesterov27method, polyak1964some, kingma2014adam, NEURIPS2020_d9d4f495}. Amongst them, some notable algorithms are Nesterov's accelerated gradient-descent (NAG)~\cite{nesterov27method}, heavy-ball method (HBM)~\cite{polyak1964some}, and Adabelief~\cite{NEURIPS2020_d9d4f495}. The above momentum-based methods improve upon the convergence rate of GD. In particular, the recent Adabelief method has been demonstrated to compare favorably for machine learning problems~\cite{NEURIPS2020_d9d4f495}. For empirical risk minimization problems with $n$ data points, the per-iteration computational cost of these algorithms is $\bigO(n d)$. However, for general  cost $f$, these algorithms converge at a {\em sublinear} rate~\cite{tong2019calibrating,su2014differential}. For the special case of strongly convex cost $f$, the aforementioned methods converge {\em linearly}~\cite{tong2019calibrating,su2014differential}.

Newton's method~\cite{kelley1999iterative} explores second-order information of $f$. Specifically, when $f$ is strongly convex, Newton's method pre-multiplies the gradient with the inverse  Hessian matrix at every iteration, resulting in local {\em quadratic} convergence rate~\cite{kelley1999iterative}. Despite of faster convergence rate, there are several issues in Newton's method. (i) For empirical risk minimization, the per-iteration computational cost of Newton's is $\bigO(n d^2 + d^3)$. (ii) Secondly, convergence of Newton's method is guaranteed only if $f$ is strongly convex and the convergence is local. (iii) Additionally, it involves computing a matrix inverse at every iteration, which is highly unstable against {\em process noise}, such as hardware failures and quantization errors. 
On the other hand, the IPG method applies to non-convex cost  and is robust against noise. A brief overview of the method is provided next.

The IPG algorithm follows the basic prototype of the gradient-descent method. However, a notable difference is that the gradients are multiplied by a {\em pre-conditioner} matrix in the IPG algorithm. The {\em pre-conditioned} gradient updates the current estimate. This technique is commonly known as {\em pre-conditioning}~\cite{nocedal2006numerical}.

Next, we describe the proposed Iteratively Pre-conditioned Gradient-descent (IPG) algorithm. The algorithm is iterative where in each iteration $t \in \{0, \, 1, \ldots\}$, an estimate $x_t \in \R^d$ of a minimum point Eq.~\eqref{eqn:opt_1} and a pre-conditioner matrix $K_t \in \R^{d \times d}$ are maintained, and updated using steps presented below.

{\bf Initialization:} Before starting the iterations, an initial estimate $x_0$ and a pre-conditioner matrix $K_0$ is chosen from $\R^d$ and $\R^{d \times d}$, respectively. Further, three sequences of non-negative scalar parameters $\{\alpha_t, \beta_t, \delta_t, t\geq 0\}$ are chosen, such that $\delta_t \leq 1$, $\beta_t > - \lambda_{\min} (H_t)$, and $\alpha_t < \frac{1}{\lambda_{\max} (H_t) + \beta_t}$. Here, $\lambda_{\min} (\cdot)$ and $\lambda_{\max} (\cdot)$ respectively denote the smallest and the largest eigenvalue of a square matrix.



{\bf Steps in each iteration $\bm{t}$}: For each iteration $t \geq 0$, we let $f_t = f(x_t)$, $g_t = \nabla f(x_t)$, and $H_t = \nabla^2 f(x_t)$ respectively denote the value of the cost function $f$, its gradient vector, and the Hessian matrix evaluated at the current estimate $x_t$.
Let $I$ denote the $(d \times d)$-dimensional identity matrix.
In each iteration $t$, the algorithm comprises two steps. 
In {\em Step 1}, the estimate $x_t$ is updated to $x_{t+1}$ such that
    \begin{align}
        x_{t+1} & = x_t - \delta_t K_t g_t. \label{eqn:x_update}
    \end{align}
In {\em Step 2}, the pre-conditioner matrix $K_t$ is updated to $K_{t+1}$:
    \begin{align}
        K_{t+1} & = K_t - \alpha_t \left(\left(H_t + \beta_t I\right)K_t - I\right). \label{eqn:K_update}
\end{align}

In deterministic settings, the convergence analysis of IPG algorithm can be found in~\cite{chakrabarti_accelerating_2021}. In the presence of noise, $x_t$ in the IPG algorithm is expected to converge to a neighborhood of a stationary point $x_{st}$~\cite{chakrabarti_accelerating_2021-1}. Empirically, the IPG algorithm has been implemented for solving standard convex and non-convex classical optimization problems, including binary classification on the MNIST dataset, noisy quadratic model of neural network training, and beamforming for wireless communication in contested environments, subject to process noise corrupting the iterates of the algorithm. To solve these problems, IPG requires fewer iterations and fewer floating point multiplications to reach the desired accuracy and obtains a smaller steady-state error compared to the existing gradient-based first-order optimizers and quasi-Newton optimizers such as BFGS~\cite{chakrabarti2022control}. While the faster convergence of IPG is attributed to the pre-conditioner $K_t$ enabling the iterations~\eqref{eqn:x_update}-\eqref{eqn:K_update} to asymptotically converge to Newton's method, the improved robustness against noise is attributed to the asymmetry and non-positive definiteness of $K_t$ which is in contrast with other fast converging quasi-Newton optimizers such as BFGS.

\section{Quantum circuit optimization using IPG}

The problem of quantum circuit optimization consists of converting a quantum algorithm into a set of realizable gates and measurements. While some of these gates could be hardware-specific, DiVincenzo's criteria lays down general rules that nearly all universal quantum computing platforms should satisfy \cite{divincenzo_physical_2000}. Two of these criteria are the ability to implement a universal gate set \cite{deutsch_universality_1995} and the ability to initialize the qubit set into a known initial state (often considered the state where all qubits are initialized to $|0\rangle$). Here we take the problem of quantum circuit optimization to be the construction of an efficient sequence of gates chosen from this universal gate set, which consists of single-qubit gates and the two-qubit controlled NOT gate in our case. The final optimized circuit approximates the unitary transformation required to implement a quantum algorithm to a high fidelity, close to unity. Note that this differs from variational quantum algorithms and quantum optimization algorithms that usually optimize for the expectation value of an operator computed from multiple shots \cite{muller_accelerating_2022, verdon_learning_2019, ostaszewski_reinforcement_2021, radin_classically-boosted_2021}. Our approach aims to mimic the entire quantum operator needed to implement an algorithmic subroutine or to prepare a target quantum state, such as a maximally entangled (Greenberger-Horne-Zeilinger) GHZ state.

The importance of this problem for progress in quantum computing is underscored by current limitations in the number of qubits permitted by hardware realizations, be it trapped ions, superconducting qubits, neutral atoms, photonics, or other platforms. Moreover, since each gate operating on a physical qubit is imperfect, the overall circuit error increases with increased gates. As mentioned earlier, this problem has been previously approached using first-order gradient-based methods \cite{fosel_quantum_2021, ostaszewski_reinforcement_2021, verdon_learning_2019, muller_accelerating_2022, bartlett_universal_2020, bartlett_deterministic_2021, da_silva_global_2013, shi_quantum_2022}.

\begin{figure}
    \centering
    \includegraphics[width=0.45\textwidth]{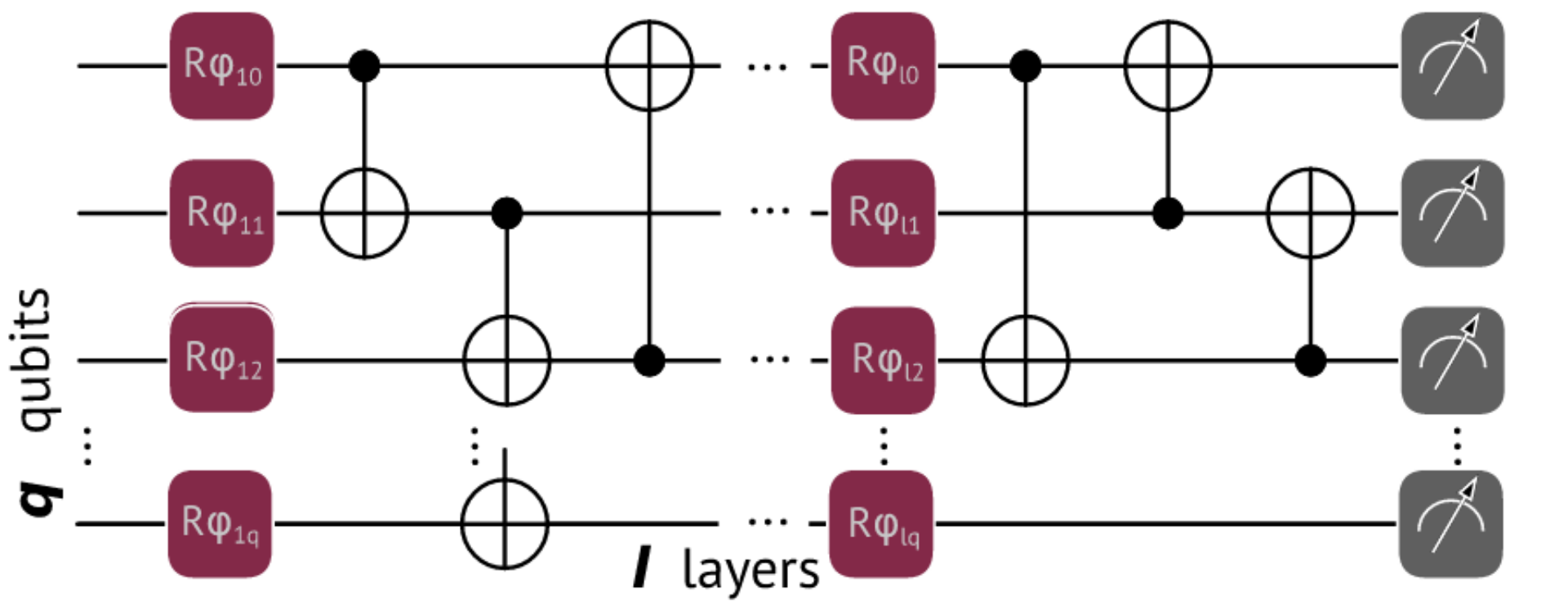}
    \caption{A representative schematic of a circuit comprising several single qubit gates labeled by $R_\phi$ and two-qubit controlled NOT gates acting on $q$ input qubits, grouped into $\ell$ layers, to implement a desired unitary transformation for a quantum algorithmic step. Quantum circuit optimization involves finding optimal parameters $\phi$ for the single qubit gates to approximate the desired circuit. In practice, we choose a template consisting of three single qubit gates $R_z(\phi_\alpha), R_y(\phi_\beta)$ and $R_z(\phi_\gamma)$ for each of the $q$ qubits per layer of the circuit. For example, $R\phi_{12} = R_z(\phi_{\alpha 12}) R_y(\phi_{\beta 12}) R_z(\phi_{\gamma 12})$.}
    \label{fig:circuit_sample}
\end{figure}

Higher-order gradient descent methods, typically belonging to the class of Newton and quasi-Newton methods, use second-order information through the Hessian matrix (see Section II) in addition to the first-order gradient vector. Hence it is important to evaluate the gradient vectors and Hessian matrices for each optimization step as accurately as possible. Finite difference methods to estimate the gradient or parameter shift methods, could both be costly on quantum hardware and could lead to imprecision in the evolution of gradients, which would propagate even more strongly to the Hessian.

To avoid such issues, we take advantage of automatic differentiation, which can compute the exact derivative of a scalar function with respect to its input parameters up to machine precision by recursively applying the chain rule \cite{paszke_automatic_2017}. It is known that the reverse mode of automatic differentiation propagates the derivatives through computational graphs or ``tapes" that record intermediate values and their dependencies, making it faster than both symbolic differentiation and numerical derivatives using finite differences \cite{griewank_automatic_1989, griewank_evaluating_2008}.

\section{Proposed Evaluation and Simulation Results}
In this section, we describe our construction of the quantum circuit and the details of its optimization. Our circuit implementation follows a template consisting of several parameterized single-qubit gates $R_\phi$ and non-parameterized two-qubit CNOT gates, as shown in Fig.~\ref{fig:circuit_sample}, acting on an input state of $q$ qubits. The circuit is grouped into $\ell$ layers. For each randomly initialized run of optimization, the number of layers $\ell$ and the number of qubits $q$ is kept fixed. Each layer \{1, 2, ... $\ell$\} consists of three single-qubit gates per qubit parameterized by angles $\phi_\alpha, \phi_\beta$ and $\phi_\gamma$ such that $R_\phi = R_z(\phi_\alpha)\, R_y(\phi_\beta)\, R_z(\phi_\gamma)$ represents the arbitrary single-qubit rotation experienced by the qubit up to a global phase. Two-qubit entangling CNOT gates combined with these single-qubit gates allow for the implementation of universal operations on the input qubits. The number of layers $\ell$ typically depends on the complexity of the quantum operation; for example, converting a separable input quantum state, say $|0\rangle^{\otimes N}$, into a highly entangled state requires propagation through a larger number of layers than the construction of quantum states with minimal entanglement. During the optimization routine, if a high fidelity was not obtained in several runs $N\sim q$ starting from different random initialization of the circuit, $\ell$ was increased in units of 2 till a desired fidelity was approached.

In our construction of the quantum circuit, the number of parameters optimized is $N_p = 3q\ell$, which grows linearly in both the number of qubits and the circuit depth. As discussed in the last section, we use automatic differentiation to calculate the gradients and Hessians that feed into the IPG optimizer. For numerical evaluation of quantum circuits, we use the python package pennylane, as it provides a differentiable quantum programming interface. We tested our code using both the Pytorch interface and the autograd interface of the pennylane package, both showing similar relative speedups. However, the Pytorch interface is faster at evaluating automatic gradients.

\subsection{State vector optimization}

\begin{figure}
    \centering
    \includegraphics[width=0.5\textwidth]{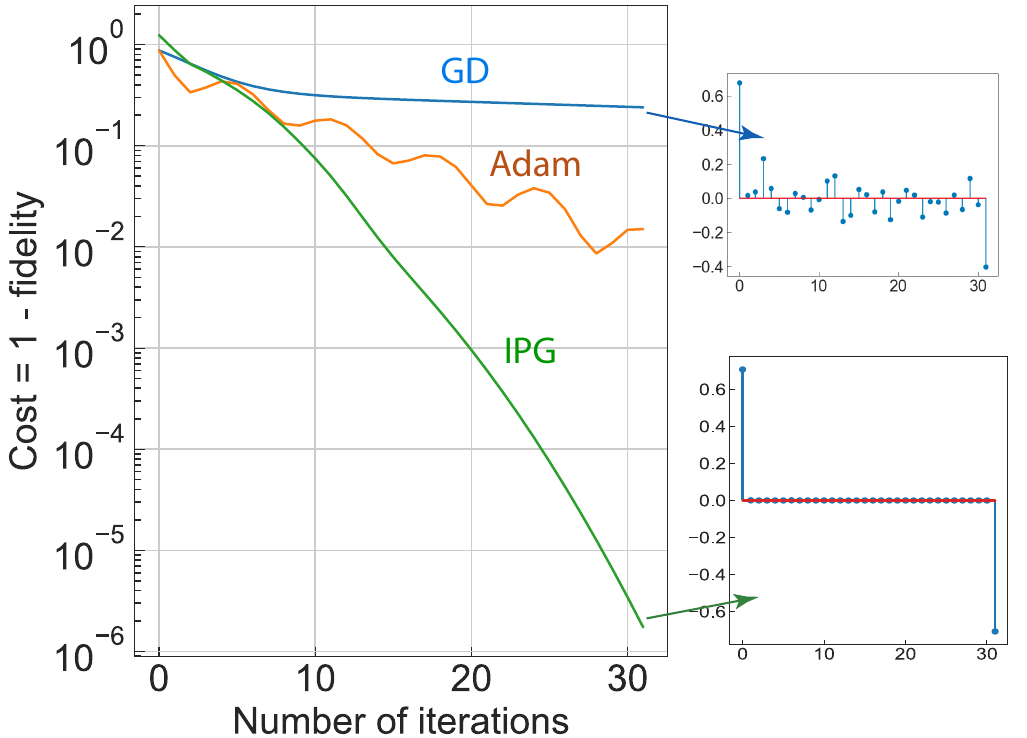}
    \caption{Three-layer quantum circuit optimization of a 5-qubit GHZ state $(|00000\rangle - |11111\rangle)/\sqrt{2}$. Left: Cost history over 32 iterations for traditional gradient descent (GD), Adam, and iteratively preconditioned gradient descent (IPG). The initial input state is chosen to be a vacuum state $|00000\rangle$. Right insets: IPG approaches the GHZ state to a high fidelity $F = 1-10^{-6}$, whereas the GD output has substantial undesired components at states other than $|00000\rangle $ and $|11111\rangle$, resulting in low fidelity $F<0.7$. The $x$ axis in the insets is labeled using the decimal equivalent of the binary qubit string, such that 0 and 31 correspond to $|00000\rangle$ and $|11111\rangle$ respectively.}
    \label{fig:performance_state_prep}
\end{figure}

We demonstrate the optimization of a quantum circuit using traditional GD and the proposed IPG method for preparing a maximally entangled GHZ state in Fig.~\ref{fig:performance_state_prep}.
The GHZ state for $N$ qubits:
\begin{equation}
|\Psi_{\rm GHZ\pm}\rangle = (|0\rangle^{\otimes N} \pm |1\rangle^{\otimes N})/\sqrt{2}
\end{equation}
is a quantum state with strong correlations between multiple qubits. It is commonly used as a benchmark to prototype the capability of quantum hardware and software since it can be difficult to prepare due to the maximal degree of entanglement in the state, which needs to be generated from the initial state, which is typically unentangled. The results in Fig. \ref{fig:performance_state_prep} were obtained  for $q=5$ qubits. We started with a different set of random initial phases of the single-qubit gates from the uniform distribution $\phi_j \sim {\rm Uniform}(-\pi, \pi)$ and chose the best of three runs of gradient descent. 
The fidelity between the target GHZ state and the output of the quantum circuit $\Psi_{\rm output} = U \Psi_{\rm input}$ is defined as usual:
\begin{equation}
    \mathcal{F} = \Re \langle \Psi_{\rm GHZ-} | \Psi_{\rm output} \rangle
\end{equation}
where $U$ represents the unitary transformation performed by the quantum circuit on the input state $\Psi_{\rm input}$. We choose the real part and not the absolute value of the inner product to retain the phase information. The cost for optimization is defined as $1-\mathcal{F}$. We empirically choose a learning rate $\eta = 0.09$; larger or smaller learning rates were also tested, with similar or slightly worse performance observed. 

It is clear that our proposed IPG method outperforms both gradient-descent (GD) and Adam by four orders of magnitude after 32 iterations, converging to a near-unity fidelity to better than $10^{-6}$, while GD and Adam approach infidelity of 0.5 and $10^{-2}$ respectively. $N_p=45$ single-qubit gate rotation angles were optimized for the 3-layer 5-qubit circuit. The insets in the right show the optimized state obtained by GD and IPG, with the IPG output closely resembling the expected GHZ state, whereas the GD output contains spurious components at states other than $|0\rangle^{\otimes 5}$ and $|1\rangle^{\otimes 5}$.

We next discuss the circuit optimization results to prepare another class of entangled states, the $N$-qubit W state:
\begin{equation}
|\Psi_{\rm W}\rangle = \frac{|00...001\rangle + |00..010\rangle + |00...100\rangle + \cdots |10...000\rangle}{\sqrt{N}} \label{eq:W}
\end{equation}
The W state exhibits entanglement but of a qualitatively different class than the GHZ state, and the two cannot be converted using local unitary operations. For instance, if one of the qubits in the GHZ state is destroyed, the entire state collapses and becomes unentangled, whereas if one of the qubits in the W state is destroyed or traced out, the remaining state still exhibits entanglement. Hence, W state preparation could test a different set of hardware capabilities and optimizer software characteristics than GHZ state preparation. 

\begin{figure}
    \centering
    \includegraphics[width=0.5\textwidth]{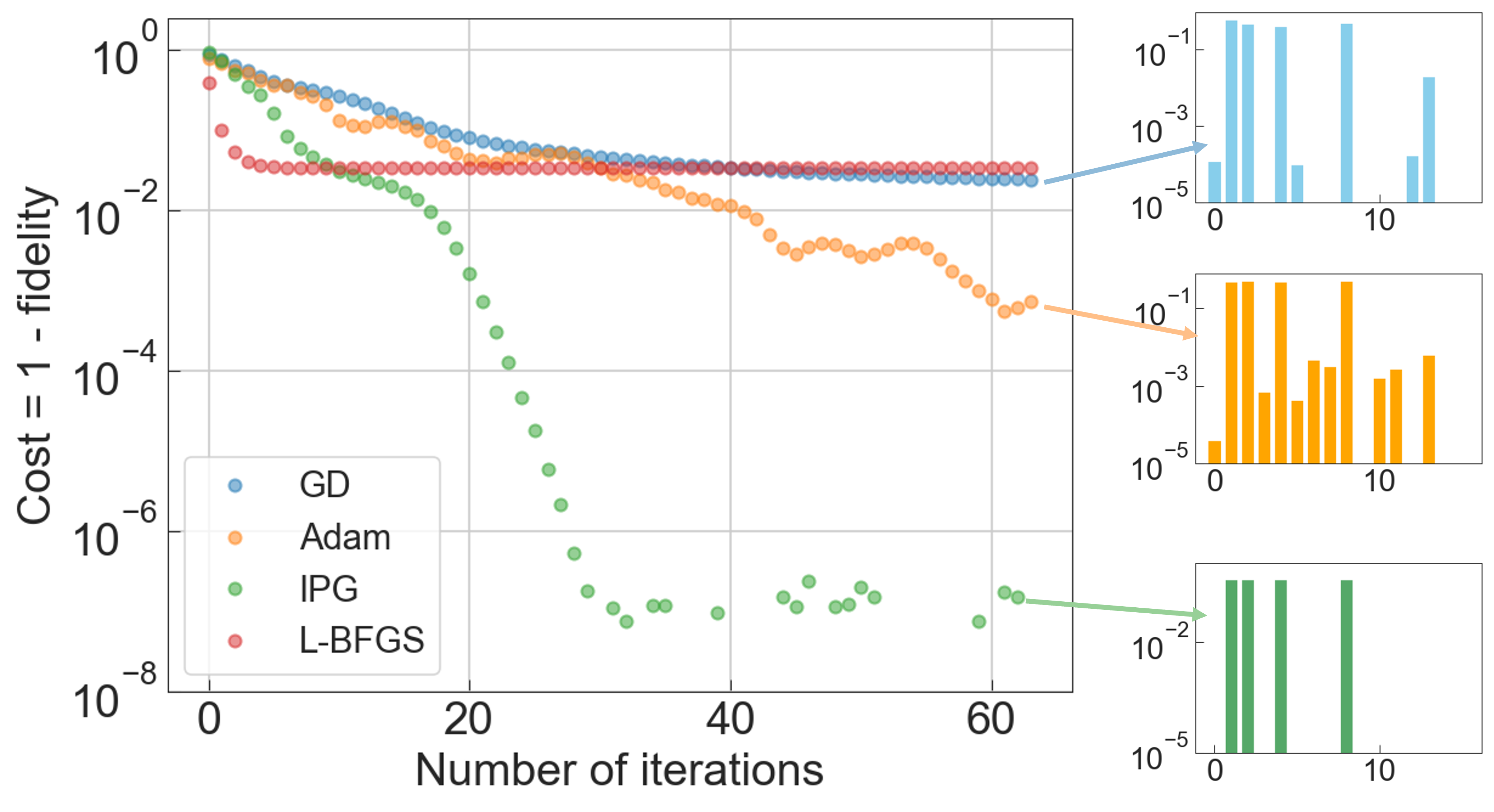}
    \caption{Results of a three-layer quantum circuit optimized to prepare the entangled state $|\Psi_{\rm W}\rangle$ of Eq. \eqref{eq:W}. 64 iteration steps for $q=4$ qubits are chosen as this is a harder optimization problem than the GHZ state preparation. IPG achieves a high-fidelity circuit, approaching fidelity of $F=1-10^{-7}$ by optimizing $N_p = 36$ single-qubit rotation angles. The insets on the right are plotted in log scale, showing the amplitudes of the W-state components. The result of IPG (bottom right) shows nonzero amplitudes only at qubit strings corresponding to 1, 2, 4 and 8, as expected for a 4-qubit W state. GD and Adam show undesired nonzero amplitudes also at other state vectors. The $x$ axis in the insets is labeled using the decimal equivalent of the binary qubit string, such that 0 and 15 correspond to $|0000\rangle$ and $|1111\rangle$ respectively.}
    \label{fig:my_label}
\end{figure}

Since preparing a W state turns out to be a more challenging optimization problem than the GHZ state preparation, we increase the number of iterations to 64, reduce the number of qubits from $q=5$ to 4, reduce the number of optimization runs to 2,  and include another second-order optimization technique: limited-memory Broyden-Fletcher-Goldfarb-Shanno (BFGS) mechanism \cite{nocedal2006numerical}. While BFGS performs better than first-order optimization techniques, in this situation we commonly observe that L-BFGS stagnates in a local minimum of the cost function, while Adam and IPG are able to escape it and approach the global optimum. The learning rate for each algorithm in Fig. \ref{fig:my_label} was tuned as a hyperparameter to attain the minimum cost or infidelity at the end of 64 iterations of optimization. Note that in several cases L-BFGS was observed to converge at a rate comparable to IPG; however, L-BFGS was observed to stagnate in a local minimum more predominantly than IPG over several random gate initializations. Some of the authors have previously reported the benefits of the preconditioner matrix in IPG to converge better when the problem is ill-conditioned, and we anticipate that the situations where IPG performs better than L-BFGS in addition to the first-order algorithms could be one such example. Thus we have empirically illustrated that IPG offers performance advantages in terms of converging to a high fidelity circuit with $F \approx 1-10^{-7}$ for W state preparation.

\begin{figure}
    \centering
    \includegraphics[width=0.44\textwidth]{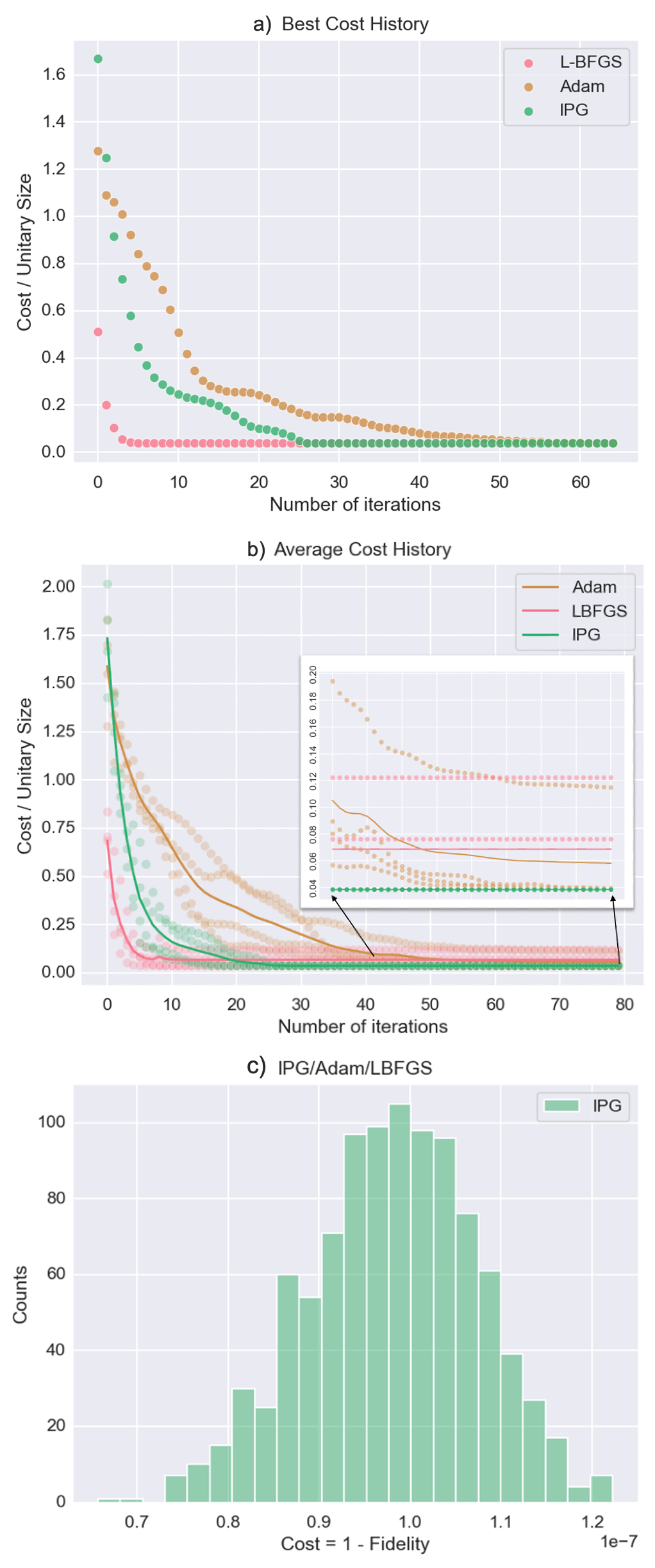}
    \caption{Three-qubit Quantum Fourier Transform (QFT) optimization using the matrix distance length. \textbf{a.} \textit{Best} cost history over 4 randomly initialized runs of 80 iterations for L-BFGS, Adam, and IPG. All optimizers arrive at a Cost/Unitary Size value of 0.0384 at varying iterations of convergence. \textbf{b.} \textit{Average} Cost/Unitary Size over 4 runs and 80 iterations for L-BFGS, Adam, and IPG. IPG remains consistent in performance each run, with the number of iterations and final Cost/Unitary Size value remaining consistent each run. This contrasts L-BFGS and Adam, which vary both by degree of convergence and number of iterations needed to converge at each run's lowest Cost/Unitary Size. \textbf{c.} Histogram of fidelity between the output state of the optimized parametrized circuit, and the output state of an ideal QFT unitary. 1000 random input states were generated and applied to both unitaries. Results show good infidelity or cost of $10^{-7}$.
    }
    \label{fig:qft}
\end{figure}

\subsection{Full circuit unitary optimization}

We evaluate the performance of the IPG method for optimizing the full unitary matrix corresponding to a quantum circuit, instead of the quantum state preparation routines discussed above. Unsurprisingly, achieving a $2^q \times 2^q$ matrix representing the circuit turns out to be more challenging than preparing a $2^q$-element column vector representing a quantum state. For our numerical demonstration, we choose the quantum Fourier transform (QFT) circuit, which is widely used in quantum algorithms exhibiting exponential speedup. In fact, Shor's factoring algorithm and quantum phase estimation both use QFT. A QFT circuit acting on an input state vector $|\Psi_{\rm input}\rangle =\sum_{k=0}^{2^q-1} x_k |k\rangle $ produces an output state $|\Psi_{\rm output}\rangle = \sum_{k=0}^{2^q-1} y_k |k\rangle$ according to,
\begin{equation}
    y_k =\frac{1}{2^{q/2}}\, \sum_{j=0}^{2^q-1} x_j \omega_{2^q}^{jk} 
\end{equation}
with $\omega_{2^q} = \exp(i2\pi/2^q) $ being a complex root of unity.

The fidelity with which a general quantum circuit's full matrix is approximated was quantified using two separate methods. First, the matrix distance was evaluated, which corresponds to the average element-wise distance between the target matrix $[A]_{mn}$ and the matrix representation $[B]_{mn}$ of the optimized quantum circuit: $D = \sum_{mn} | A_{mn} - B_{mn}| ^2 /2^q$. The matrix distance $D$ was minimized using existing GD and IPG algorithms for comparison [Fig. \ref{fig:qft}(a), (b)]. Second, the overlap of the output state produced by the optimized circuit and the ideal output state produced by the target QFT circuit was calculated for 1000 randomly chosen input states. The absolute value of this fidelity was then used to assess the quality of approximating the quantum circuit by the optimized gate ansatz \ref{fig:qft}(c). Conventional gradient descent was not evaluated here due to its performance comparisons in Fig. \ref{fig:performance_state_prep}. IPG, Adam and L-BFGS were all observed to converge to small values of the average cost or matrix distance $D$ in the best-case cost history among 3 runs of random initialization [Fig. \ref{fig:qft}(a)] using a 5-layer ansatz. However, IPG was observed to converge to a lower cost in the average case, as seen in the inset of Fig. \ref{fig:qft}(b). This is because, similar to the case of W-state preparation, L-BFGS was observed to get trapped in a local minimum. Adam displayed convergence beyond the local minimum where L-BFGS was trapped, but at a slower rate. Fig. \ref{fig:qft}(c) shows the results of the histogram evaluation of output state fidelity with the expected ideal QFT output state over 1000 random input states, showing that the circuit approximates the QFT unitary to a high fidelity of $1-10^{-7}$ when optimized with IPG.

\begin{figure}
    \centering
    \includegraphics[width=0.48\textwidth]{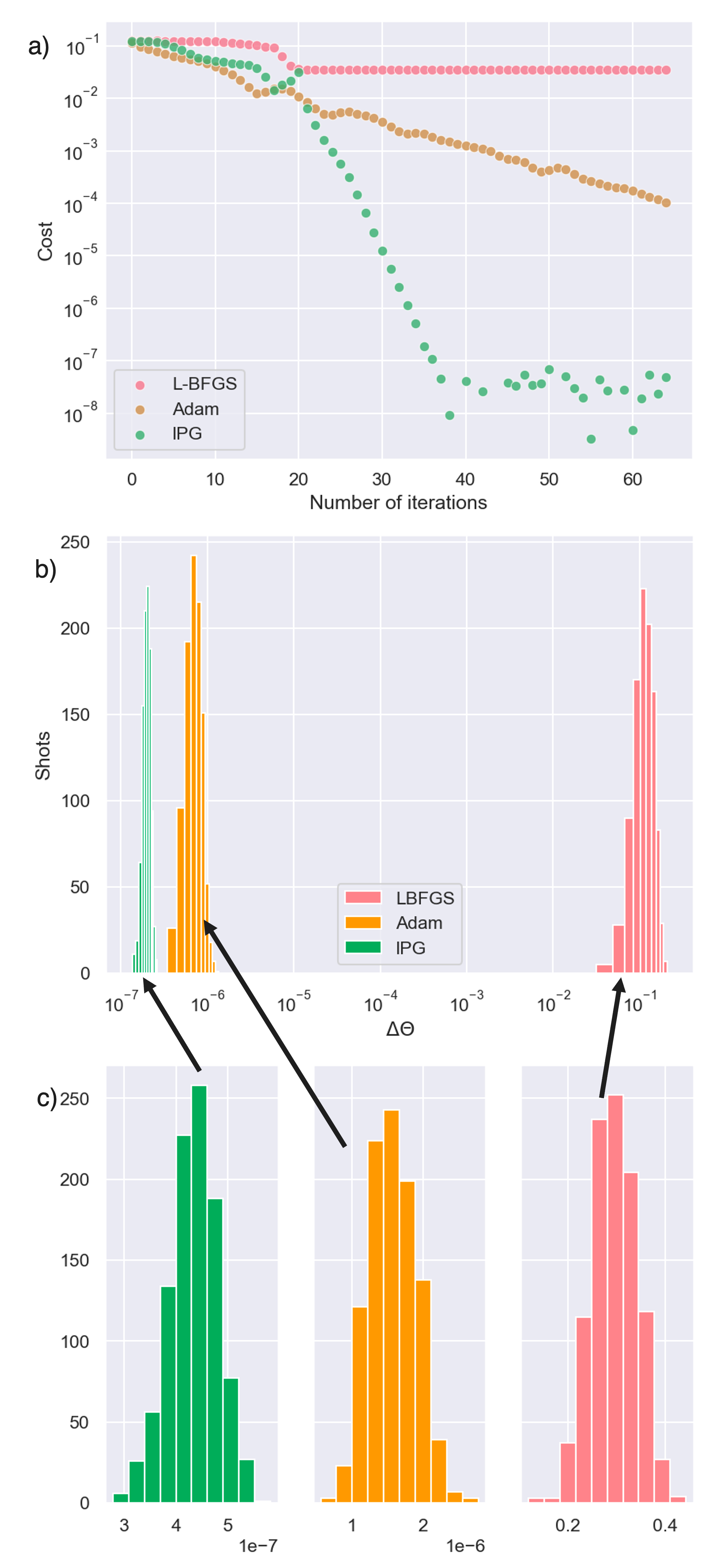}
    \caption{Three-qubit Quantum Fourier Transform (QFT) optimization using the Frobenius inner product. 
    \textbf{a.} \textit{Best} cost history over 4 runs of 64 iterations for L-BFGS, Adam, and IPG with random initializations of single-qubit gate phases. IPG arrives at cost of $2.37*10^{-7}$, approximately three orders of magnitude better than next best cost from Adam. L-BFGS trapped in local minima preventing convergence up to a global phase. \textbf{b.} 1000 random initialization states given to parameterized circuits from IPG/Adam/L-BFGS on a noiseless statevector simulator. For each initialization and optimizer, we calculate ${\Delta \theta = |\theta_{1}-\theta_{2} |}$, where 
    $ (\theta_1, \theta_2) = (\cos^{-1} \Re \langle \Psi_{\rm QFT}| \Psi_{\rm opt} \rangle, \sin^{-1} \Im \langle \Psi_{\rm QFT}| \Psi_{\rm opt} \rangle$. $\ \Delta \theta =0$ indicates perfect convergence to the QFT up to a global phase. 
    \textbf{c.} Zoomed-in log-scale plots of $\Delta\theta$ for IPG and Adam, which are able to show high-fidelity QFT up to a global phase, with a one-order of magnitude better fidelity from IPG. 
    }
    \label{fig:newcostfunc}
\end{figure}

The cost function based on the matrix distance used in Fig. \ref{fig:qft} is sensitive to a global phase offset between the target unitary and the optimized quantum circuit unitary. To evaluate the performance of the IPG algorithm using a cost function that is independent of the global phase offset, we also perform gradient descent using the matrix inner product defined as \cite{buddhiraju_arbitrary_2021, giamougiannis_coherent_2023, van_der_meer_experimental_2023},
\begin{equation}
    \mathcal{F} = \frac{\left( \sum_{ij} A_{ij}^* B_{ij}\right)^2}{\left( \sum_{ij}|A_{ij}|^2 \right)\left( \sum_{ij}|B_{ij}|^2  \right)} = \left( \sum_{ij} A_{ij}^* B_{ij}/ 2^{q}\right)^2 
\end{equation}
where the last step follows from the fact that both $A$ and $B$ are unitary matrices of size $2^q \times 2^q$. For the number of layers used here $\ell = 5$, the final optimized fidelity is significantly higher (lower infidelity) using the above definition for all gradient descent methods, see Fig. \ref{fig:newcostfunc}, with the optimal infidelity of IPG approaching $10^{-8}$. In this scenario, we also observe that L-BFGS gets saturated at a local minimum, while Adam shows convergence at a slower rate than IPG.

\subsection{Results from noisy QPU runs}
\begin{figure}
    \centering
    \includegraphics[width=0.40\textwidth]{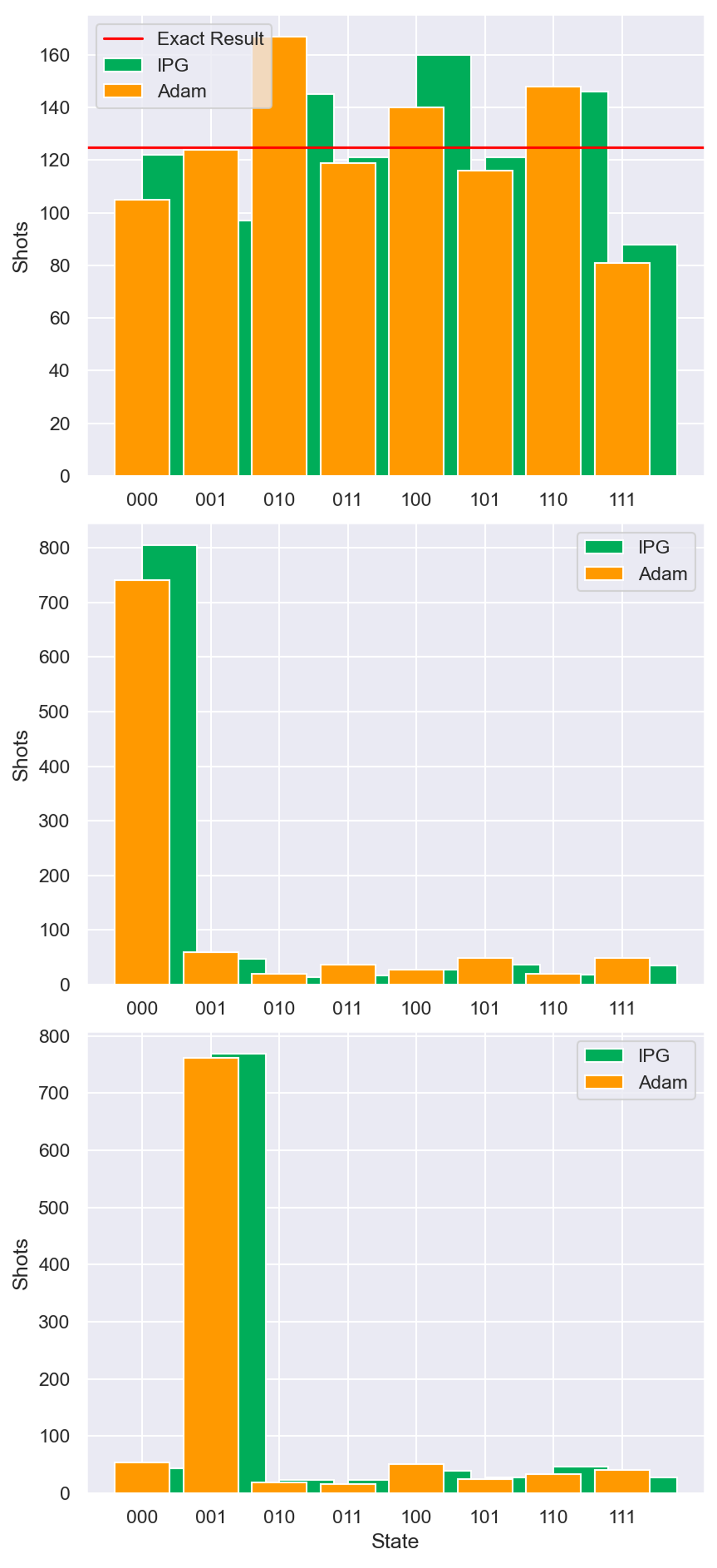}
    \caption{Quantum Processing Unit (QPU) runs on IonQ's Aria-1 simulator and hardware device for a 3-qubit quantum Fourier transform circuit. States are labeled along the x-axis based on qubit strings. {\bf a.} Output state counts for a QPU run on Aria-1 hardware device for quantum circuits optimized by Adam and IPG. The input state is $|000\rangle$. {\bf b.} Output state counts of the Aria-1 noisy simulator for quantum circuits optimized by Adam and IPG. The input state is an even superposition of all input basis states with the same amplitude and phase: $\sum_i |i\rangle /2^{3/2}$. {\bf c.} Same as b, but with an odd superposition of all input basis states. 
    }
    \label{fig:qpu}
\end{figure}

Finally, we compare the performance of the quantum circuits with optimized paramters on IonQ's trapped-ion quantum computers, using both the noisy simulator and the hardware device of the quantum processing unit (QPU). For the QPU device execution, we tested with both Harmony and Aria-1 quantum computers, the former being more noisy than the latter and hence yielding worse results as expected. Here we exclusively report results from the Aria-1 noisy simulator and QPU device. Fig.\ref{fig:qpu}(a) reports the results for a 3-qubit input state $|000\rangle$, for which the QFT output is expected to show a uniform distribution across all the $2^3$ output states. Figs. \ref{fig:qpu}(b) and (c) on the other hand show Aria-1 QPU simulator results for an input state that is in an even or odd superposition of all qubits, respectively. These states were generated from an initial $|000\rangle$ state by applying a Hadamard gate to all qubits. An additional $Z$ gate was applied to a single qubit. Note that the amplitude of this superposition is uniform, while the phase is either uniform or periodic, which manifests in the QFT output as a peak at the $|000\rangle$ or $|001\rangle$ state. Although the results show reasonable agreement with the expected outputs, we anticipate that error mitigation techniques would be necessary for future work to realize more high-fidelity outputs on the noisy QPU to provide better benchmarking of the optimized circuits.

\section{Conclusions and Discussion}

We have investigated the use of iteratively preconditioned gradient descent (IPG) as a promising algorithm for quantum circuit optimization, expanding its use case beyond those of previous classical scenarios. The optimization converges faster than traditional first-order gradient descent methods including Adam for preparing highly entangled states of several qubits such as the GHZ state and the W state. Optimization results were also compared for a full quantum circuit unitary, the quantum Fourier transform (QFT). The optimized QFT circuit was implemented on IonQ's quantum processing unit (QPU), with the output states agreeing reasonably with expected results. We anticipate the IPG technique to provide advantages especially in situations where the condition number is large, which is commonly encountered in quantum circuit optimization problems as the number of non-unique solutions is large.Other methods of quantum circuit optimization that have been recently reported could benefit from the performance gains of IPG shown here \cite{sarkar_scalable_2023, ibrahim_crosstalk-based_2023, clark_novel_2022, schmitz_optimization_2023, paykin_pcoast_2023, verchere_optimizing_2023}. Future work could include the use of distributed techniques, and thorough investigation of the IPG method in the presence of noise in the gates as well as noise in the gradient update steps, which is beyond the scope of the current work. 

The results presented in this paper are hardware agnostic and should apply to most gate-based universal quantum computing platforms. Future enhancements could look at more hardware-specific versions which take into account the preferred universal gate set for a certain platform, such as a superconducting qubit platform (IBM, Google, Rigetti) or a trapped ion platform (Quantinuum, IonQ). For example, while we have currently optimized circuit fidelities by parameterizing the single-qubit rotation gate angles without restrictions on how large the angle can be, it would be interesting to optimize the quantum circuit using the preferred range of angles for a qubit hardware platform since gates using smaller angles ($\sim\pi/100$) have shown fidelity improvement in recent experiments \cite{wright_benchmarking_2019, nam_ground-state_2020}. Alternatively, some platforms benefit by choosing from a finite discrete set of angles. Similarly, using a hardware-native basis set of gates, such as the one-qubit set of GPi and GPi2 gates, and the two-qubit set of XX, YY and XY gates, or the partially-entangling/ arbitrary phase M{\o}lmer-S{\o}rensen gate in the trapped ion platform, could be beneficial \cite{noauthor_getting_nodate}. More generally, with these constraints in mind, it would be of prime importance to explore the use of advanced gradient descent techniques to reduce costs in terms of hardware resources and speed of convergence in a variety of hybrid quantum-classical optimization scenarios.

\section*{Acknowledgments}
This work was supported by the National Quantum Lab (QLab) jointly between IonQ and the University of Maryland. We acknowledge discussions with Franz Klein and John Sawyer from the Q-Lab and Pedro Rivero from IBM Quantum.

\bibliographystyle{ieeetr}

\bibliography{My_Library, new_refs_zotero, refs_ipg}

\end{document}